# Some inequalities in the fidelity approach to phase transitions

N. S. Tonchev

Institute of Solid State Physics, Bulgarian Academy of Science, 72 Tzarigradsko Chaussee Blvd., 1784 Sofia, Bulgaria

and

J. G. Brankov

Bogoliubov Laboratory of Theoretical Physics, Joint Institute for Nuclear Research, 141980 Dubna, Russian Federation,

Institute of Mechanics, Bulgarian Academy of Sciences, 4 Acad. G. Bonchev St., 1113 Sofia, Bulgaria

Abstract: We present some aspects of the fidelity approach to phase transitions based on lower and upper bounds on the fidelity susceptibility that are expressed in terms of thermodynamic quantities. Both commutative and non commutative cases are considered. In the commutative case, in addition, a relation between the fidelity and the nonequilibrium work done on the system in a process from an equilibrium initial state to an equilibrium final state has been obtained by using the Jarzynski equality.

*Keywords*: Fidelity, fidelity susceptibility, phase transitions, Jarzynski equation

## 1. Introduction

In the last decade the study of theoretical information properties of quantum models [1] provides an alternative paradigm for understanding the critical phenomena [2]. It sheds light on the problems from a different point of view, when the standard Landau-Ginzburg approach based on the ideas of an order parameter and symmetry breaking is hampered. A prominent example is the case when topologically ordered phases, or Berezinskii-Kosterlitz-Thouless phase transitions appear, see, e.g. [3,4]. In this approach the plausible fact that the properties of different macroscopic phases of matter should be encoded in the structure of rather distinct quantum states, both pure and mixed, has been put forward. Due to its geometric meaning, the problem of similarity (closeness) between states can be readily translated in the language of information geometry. On this route, the generic quantity known as fidelity may be used to play a role similar to an order parameter. In the quantum communication theory, the fidelity is a quantitative measure of the accuracy of transmission for any given communication scheme. Being a measure of the similarity between quantum states, fidelity should change abruptly at a critical point, thus locating and characterizing the phase transition. This quantity was introduced by Uhlmann [5] as a finite-temperature functional of two density matrices, $\rho_1 \equiv \rho(h_1)$ and $\rho_2 \equiv \rho(h_2)$,

$$F(\rho_1, \rho_2) = Tr\sqrt{\rho_1^{1/2} \rho_2 \rho_1^{1/2}}, \qquad (1)$$

and called fidelity by Jozsa [6], who studied its basic properties in the context of finite dimensional Hilbert spaces (see also [2] and references therein). For concreteness and simplicity, here and below we will consider one-parameter family of Gibbs states

$$\rho(h) = [Z(h)]^{-1} \exp[-\beta H(h)],$$

defined on the family of Hamiltonians of the form $H(h) = T - hS$, where the Hermitian operators $T$ and $S$ do not commute in the general case, $h$ is a real parameter, and $Z(h) = Tr \exp[-\beta H(h)]$ is the corresponding partition function. Fidelity itself is not a distance, but closely related to it is the Bures distance

$$d_B(\rho_1, \rho_2) = \sqrt{2 - 2F(\rho_1, \rho_2)}, \qquad (2)$$

which is a measure of the statistical distance between the two density matrices $\rho_1$ and $\rho_2$. The Bures distance (2) has the important properties of being Riemannian and monotone metric on the space formed by the family of density matrices. However, because of mathematical difficulties, it is not an easy task to evaluate the fidelity analytically. As a further simplification, the concept of fidelity susceptibility (or the second derivative of the fidelity) naturally appears as a tool which may be employed in the analytical study of phase transitions. This is due to the amazing fact that fidelity susceptibility may be related to (or estimated by) some more conventional physical quantities, such as imaginary-time dynamical responses, and reveals signatures of a phase transition [2]. The fidelity susceptibility $\chi_F(\rho_0)$ arises as a leading-order term in the expansion of the fidelity for two infinitesimally close density matrices, e.g., $\rho_1 = \rho_0 - \delta\rho/2$ and $\rho_2 = \rho_0 + \delta\rho/2$. Note that the fidelity and fidelity susceptibility under consideration are defined with respect to the parameter $h$, including the important symmetry breaking case when the system undergoes a phase transition as $h$ is varied. The fidelity susceptibility at the point $h = h_0$ in the parameter space is conveniently defined (in a symmetric form) as

$$\chi_F(\rho(h_0)) :=$$
$$\lim_{\delta h \to 0} \frac{-2 \ln F(\rho(h_0 - \delta h/2), \rho(h_0 + \delta h/2))}{(\delta h)^2} \quad (3)$$

From the above definitions, we obtain for the case of two infinitesimally close density matrices the following relation between the Bures distance and the fidelity susceptibility:

$$d_B^2(\rho(h_0 - \delta h/2), \rho(h_0 + \delta h/2)) =$$
$$\chi_F(\rho(h_0))(\delta h/2)^2 + O((\delta h/2)^4), \quad \delta h \to 0.$$

The quantity $\chi_F(\rho(h_0))$ is more convenient for studying than the fidelity itself, because it depends on a single point $h = h_0$ of the parametric manifold. Physically, it is a measure of the fluctuations of the driving term which is introduced in the Hamiltonian through the parameter $h$.

## 2. Commutative case

In the case when $\rho_1 \equiv \rho(h_1)$ and $\rho_2 \equiv \rho(h_2)$ commute (i.e., the operators $T$ and $S$ commute), there are simple relations between the fidelity and the partition function

$$F(\rho_1, \rho_2) = \frac{Z\left(\frac{h_1 + h_2}{2}\right)}{\sqrt{Z(h_1)Z(h_2)}}. \quad (4)$$

This relation allows one to understand the state evolution at finite temperatures from the knowledge of thermodynamics. On the other hand, if we introduce the random quantity $W_{1,2} := W(h_1, h_2)$ as the work done on the system by an outside agent in an arbitrary process from $h_1$ to $h_2$, the following equality takes place:

$$\overline{\exp(-\beta W_{1,2})} = \frac{Z(h_2)}{Z(h_1)} \quad (5)$$

The overbar indicates an ensemble average over all possible paths through phase space from $h_1$ to $h_2$. This equation was first discovered in 1997 by C. Jarzynski and came to be known as the Jarzynski equality (see, e.g., [7]). It relates the nonequilibrium average work done by a driving force on a system, initially at equilibrium, to the ratio of the partition functions for the two (initial and final) equilibrium states. Some experimental tests of this exact relation of nonequilibrium statistical mechanics, performed on meso- and nanosystems, are discussed in [8]. It is easily seen that one can relate the l.h.s. of equalities (4) and (5). First, we use the Jarzynski equality for a process from $h_1$ to $h_0$ and after that for a process from $h_2$ to $h_0$. We obtain

$$\overline{\exp(-\beta W_{10})} = \frac{Z(h_0)}{Z(h_1)} \quad \text{and}$$
$$\overline{\exp(-\beta W_{20})} = \frac{Z(h_0)}{Z(h_2)}.$$

Next, multiplying these equalities, we get

$$\overline{\exp(-\beta W_{10})} \times \overline{\exp(-\beta W_{20})} = \frac{Z^2(h_0)}{Z(h_1)Z(h_2)}. \quad (6)$$

If we choose $h_0 = \frac{h_1 + h_2}{2}$, then the r.h.s. of the above equality becomes exactly the square of the fidelity $F(\rho_1, \rho_2)$. Here we point out that in this way the related nonequilibrium quantities get involved in the considered metric approach (at least formally). The question of the definition of nonequilibrium work in the Jarzynski equality is the main topic of the existing discussion in the literature, see [9,10]. One should be able to incorporate the phase transition phenomenon in the present consideration, but here we shall not go deeper into this rather difficult issue. Let us give a hint for this intricate possibility. Considering two nearby states on the manifold of density operators $\{\rho(h)\}$ with $h_1 = h_0 - \delta h/2$ and $h_2 = h_0 + \delta h/2$ we shall introduce the fidelity susceptibility into the subject. By using the Jensen inequality $\overline{\exp(-\beta W)} \geq \exp(-\beta \overline{W})$ we obtain the inequality

$$F^2(\rho(h_0 - \delta h/2), \rho(h_0 + \delta h/2)) \geq$$
$$\exp(-\beta \overline{W_{10}}) \times \exp(-\beta \overline{W_{20}}),$$

or equivalently

$$\frac{-2 \ln F(\rho(h_0 - \delta h/2), \rho(h_0 + \delta h/2))}{\delta h^2} \leq$$
$$\frac{\beta(\overline{W_{10}} + \overline{W_{20}})}{\delta h^2}.$$

In the limit $\delta h/2 \to 0$ the l.h.s. gives noting but the symmetric fidelity susceptibility at the point $h_0$, and so

$$\chi_F(\rho(h_0)) \leq \lim_{\delta h^2 \to 0} \frac{\beta(\overline{W_{10}} + \overline{W_{20}})}{\delta h^2}, \quad (7)$$

if the corresponding limit exists. It is still questionable whether equation (6) and inequality (7) carry useful information about nonequilibrium systems without further examination. If we assume that $\overline{W_{00}} = \overline{W(h_0, h_0)} = 0$, we get

$$\chi_F(\rho(0)) \leq \frac{\beta}{4} \frac{\partial^2 \overline{W(h, h_0)}}{\partial^2 h}\bigg|_{h=h_0} \quad (8)$$

which is easily derived with the aid of the relation

$$\frac{\overline{W(h_0 - \delta h/2, h_0)} + \overline{W(h_0 + \delta h/2, h_0)}}{\delta h^2} =$$
$$\frac{1}{4} \frac{\partial^2 \overline{W(h, h_0)}}{\partial^2 h}\bigg|_{h=h_0} + O((\delta h)^4).$$

If the process is reversible and isothermal, one has $\overline{W(h, h_0)} = Nf(h_0) - Nf(h)$, where $f(h)$ and $f(h_0)$ are the equilibrium free energies densities calculated with the Hamiltonian $H(h)$. Here $N$ is the number of interacting particles (or spins). In this case the second derivative of $\overline{W(h,h_0)}$ with respect to the field $h$ is exactly the usual thermodynamic susceptibility $\chi_N(h)$. Thus, we finally obtain

$$\chi_F(\rho(0)) \le \frac{\beta N}{4}\chi_N(h_0), \qquad (9)$$

which in comparison with the directly derived equality

$$\chi_F(\rho(0)) = \frac{\beta N}{4}\chi_N(h_0) \qquad (10)$$

seems a rather trivial result. We emphasize, however, that the initial inequalities (7) and (8) may contain much stronger information if one avoids this rather restrictive treatment of the nonequilibrium part.

**3. Noncommutative case**

In the noncommutative case, when the driving term does not commute with the Hamiltonian, such type of equalities like (4) and (10) are unknown. For the problems arising in this quantum case the reader may consult the recent review [11]. An efficient means for solving the problem may provide different inequalities. It is well known that inequalities are wide-spread and traditional, as a tool for obtaining important statements, both in the theory of phase transition and in the information theory. Thereby obtained results are exact and cannot be inferred from any perturbation approach. As a step in this direction we present upper and lower bounds on the fidelity susceptibility in terms of some macroscopic thermodynamic quantities, like susceptibilities and thermal average values, which were derived in our work [12]. In order our definition for the fidelity susceptibility used in [12] to be compatible with the definition (2) and the results presented below one has to make the change $h_0 \to h_0 + \delta h/2$ in the latter one. The following expressions for the fidelity susceptibility are the basic ones for our consideration:

$$\chi_F(\rho(h_0)) = \frac{\beta^2}{4}\langle(\delta S_d)^2\rangle_0 + \frac{1}{2}\sum_{m,n,m\ne n}\left[\frac{\rho_n(h_0)-\rho_m(h_0)}{E_m-E_m}\right]^2 \frac{|\langle n|S|m\rangle|^2}{\rho_m(h_0)+\rho_n(h_0)} \qquad (11)$$

or equivalently

$$\chi_F(\rho(h_0)) = \frac{\beta^2}{4}\langle(\delta S_d)^2\rangle_0 + \frac{\beta^2}{8}\sum_{m,n,m\ne n}\frac{\rho_n(h_0)-\rho_m(h_0)}{X_{mn}}\frac{|\langle n|S|m\rangle|^2}{X_{mn}\coth X_{mn}}, \qquad (12)$$

where we assume that the Hermitian operator $H(h_0) = T - h_0 S$ has a complete orthonormal set of eigenstates $|n\rangle$, $H(h_0)|n\rangle = E_n|n\rangle$, where $n = 1,2,...$, with nondegerate spectrum $\{E_n\}$, $X_{mn} = \frac{\beta(E_m-E_n)}{2}$ and $\delta S_d = S_d - \langle S_d\rangle_0$, $S_d$ being the diagonal part of the operator $S_d$. As usual $\langle ...\rangle_0 := \sum_n \rho_n(h_0)\langle n|...|n\rangle$. In this basis the density matrix $\rho(h_0)$ is diagonal too:

$$\langle m|\rho(h_0)|n\rangle = \rho_m(h_0)\delta_{m,n}, \ m,n = 1,2,...$$

Relations (11) and (12) tell apart the classical contribution and the quantum contribution. The second term takes into account the generic noncommutativity of $\rho_1$ and $\rho_2$ in definition (1). With the help of the inequality

$$(\rho_n(h_0) - \rho_m(h_0))^2 \ge [\rho_n(h_0) + \rho_m(h_0)]\left[\sqrt{\rho_m(h_0)} - \sqrt{\rho_n(h_0)}\right]^2,$$

one obtains from eq. (11)

$$\chi_F(\rho(h_0)) \ge \chi_F^G(\rho(h_0)) + \frac{\beta^2}{8}\langle(\delta S_d)^2\rangle_0, \qquad (13)$$

where $\chi_F^G(\rho(h_0))$ is the alternative definition of the finite-temperature generalization of the fidelity susceptibility introduced in [13]. Inequality (13) is a little bit stronger version of the similar inequality obtained in [13]. With the help of the elementary inequalities $1 - \frac{1}{3}x^2 \le (x\coth x)^{-1} \le 1$, from eq. (12) one obtains

$$\frac{\beta N}{4}\chi_N(\rho(h_0)) - \frac{\beta^3}{48}\langle[[S,H(h_0)],S]\rangle_0 \le \chi_F(\rho(h_0)) \le \frac{\beta N}{4}\chi_N(\rho(h_0)) \qquad (14)$$

where $\chi_N(\rho(h_0))$ is the usual thermodynamic susceptibility defined as $-\frac{\partial^2 f[H(h_0+\delta h)]}{\partial^2(\delta h)}|_{\delta h=0}$ and $f[H(h_0+\delta h)]$ is the free energy density of the $N$-particle system described by the Hamiltonian $H(h_0+\delta h)$. Inequalities (14) imply that if the term with the double commutator in the l.h.s. is finite, a divergence in $N\chi_N$ must lead to a divergence in $\chi_F$ and vice versa.

**4. Applications to specific models and comments**

In our paper [12] some of the above obtained bounds were tested on concrete models. The quality of the bounds was checked by the exact expressions for a single spin in an external magnetic field. We also considered two examples of popular many-particle models: the Dicke superradiance model and the single impurity Kondo model. The Dicke model is commonly used to illustrate how an atomic ensemble spontaneously emits electromagnetic wave with an intensity proportional to the square of the number of atoms $N^2$ rather than to $N$, as one would expect if the atoms radiate incoherently. Recently there is a renewed interest in this issue due to studies of the existing superradiant phase transition in the context of quantum entanglement (see, e.g. the review [14] and references therein). The Kondo model believed to described a rich variety of physical phenomena, e.g. quantum phase transitions, non-Fermi liquid behavior, nonconventional superconductivity, etc., (see [15]). Recent considerations have shown that the Kondo model and its generalizations are attractive candidates for quantum information processing [16]. In both models, the calculations showed the breakdown of perturbation theory as the temperature is reduced. That is why they present exactly the cases when the inequality approach can be used in order to obtain instructive results.

Here we have shown that a Jarzynski tape relation, see (6), between the average exponentials of the corresponding thermodynamic works and the square of the fidelity takes place in the commutative case. A further implication of this equation is inequality (7) which estimates a metric quantity, the fidelity susceptibility, through nonequilibrium thermodynamic ones. In the noncommutative case, we have established bounds (14) on the fidelity susceptibility which are expressed in terms of equilibrium thermodynamic quantities. One can infer from (14) that, as far as divergent behavior is considered, the fidelity susceptibility and the thermodynamic susceptibility are equivalent for a large class of models exhibiting critical behavior. A sufficient condition for this is the term with the double commutator in (14) to be finite.